\documentclass[aps,prb,twocolumn]{revtex4-2}
\usepackage{amsmath}
\usepackage{amsfonts}
\usepackage{amssymb}
\usepackage{graphicx}
\usepackage{tabularx}
\usepackage{color}
\usepackage{chemformula}

\begin{document}
\title{Size-dependent ferroelectric-to-paraelectric {sliding transformations} and
antipolar-to-ferroelectric topological phase transitions in binary homobilayers}
\author{Alejandro Pacheco-Sanjuan,$^1$ Pradeep Kumar,$^2$ and Salvador Barraza-Lopez$^{2,3}$}
\email{alejandro.pachecos@usm.cl}
\email{pradeepk@uark.edu}
\email{sbarraza@uark.edu}
\affiliation{1. Department of Mechanical Engineering, Universidad T{\'e}cnica Federico Santa Mar{\'i}a, Valpara{\'i}so, Chile\\
2. Department of Physics, University of Arkansas, Fayetteville, AR 72701, USA\\
3. MonArk NSF Quantum Foundry, University of Arkansas, Fayetteville, AR 72701, USA}

\begin{abstract}
The recent discovery of ferroelectric behavior in few-layer materials, accompanied by the observation of antipolar domains in hexagonal boron nitride and transition metal dichalcogenide moir\'e bilayers, is paving the way for revolutionary advancements in the generation and manipulation of intrinsic electric dipoles through stacking. In addition, these cutting-edge quantum materials are reshaping our comprehension of phase transitions. Within the present study, we unveil a hitherto unreported size-dependent {sliding} behavior that marks a significant departure from conventional ferroelectrics. We also shed light on thermally induced spontaneous hyperlubric sliding within moir\'e bilayers, which can be used as a signal to distinguish topological phase transitions from an antipolar onto a ferroelectric bilayer. Our findings also suggest that the (topological) pinning of AA {nodes} in antipolar moiré homobilayers prevents the occurrence of an antipolar-to-paraelectric {transformation}.
\end{abstract}
\maketitle

\section{Introduction}

Two-dimensional (2D) ferroelectrics are fundamentally different from ternary bulk ferroelectric oxides. Those differences stem from (i) a chemistry that now includes binary \cite{KaiScience,Li2017,Yang2018,RevModPhys.93.011001} or even elemental \cite{2dbismuththeory,bismuthmonolayers,graphiteFerro,Wu2023,GraphiteMoshe} materials, (ii) the presence of new mechanisms for removing centers of inversion such as relative rotations within a homobilayer \cite{Li2017} or by the simple creation of a heterobilayer, and (iii) by the presence of ferroelectric-to-paraelectric phase transitions \cite{Liu2022} by relative sliding \cite{Bauer2DFerroelectricsSliding,Marmolejo-Tejada2022} { (whereby the putative paraelectric structure is a time average over ferroelectric configurations that change the direction of their intrinsic electric dipole at discrete sliding events)}. Bilayer ferroelectrics are created by a $60^{\circ}$ rotation of one binary monolayer on a bilayer stack, away from a ground state configuration \cite{Li2017}, and they feature a macroscopically large number of degenerate minima \cite{Marmolejo-Tejada2022}.

Despite the existence of works displaying double-well energy potentials \cite{Marmolejo-Tejada2022,Bauer2DFerroelectricsSliding} and of analytical calculations describing ferroelectric-to-paraelectric phase transitions in those homobilayers, the {\em size-dependence} of the {propensity for} sliding has not been addressed to date. Furthermore, it is difficult to realize a precise $60^{\circ}$ rotation among bilayers \cite{Lau}, and a more likely outcome is to achieve rotations by $60^{\circ}-\delta$, where $\delta$ is a small angle. As it is well-known by now, such mismatch gives rise to moir\'es, which have originally been studied for their unique electronic properties \cite{LopesdosSantos2007,Andrei2009,Didier,Morell2010,PhysRevLett.121.266401,PhysRevLett.121.026402,TMDs1,Fu2020,CorrelatedTMDC,Lede} but also display antipolar domains on bilayers made from binary compounds \cite{WTe2,MoS2,paper4,Liu2022,WuPNAS,2023Cazeaux,2022Cazeaux}. According to Bennett and coworkers, binary moir\'e homobilayers are antipolar, {\em not} antiferroelectric \cite{Bennett1,Bennett2}. {As superbly described by Cazeaux and coworkers, such antipolar structures feature AB and BA triangular domains separated by dislocation lines meeting at AA-nodes \cite{2022Cazeaux}.} Bennett and coworkers argue that the swap of polarization among AB and BA domains is tied to a topological winding of the local intrinsic dipole moment (which thus also includes in-plane components), with a meron and antimeron texture at AB and BA domains, respectively \cite{Bennett3}. Such topological behavior may be readable through superlubricity, but a direct comparison of the sliding behavior of ferroelectric and antipolar bilayers has not been presented yet.

{In addition, there should be} a coupling of such intrinsic electric dipole with non-trivial topology and the inherent topology induced by strain in moir\'e hBN bilayers \cite{paper1,paper2,paper3,paper4}, which have {one AA node per moir\'e supercell}. {\em The density of domains cannot be changed spontaneously} but only by a change of the angle of rotation, strain, or a combination of both \cite{2023Cazeaux,2022Cazeaux}. Atomistic reconstructions take place such that AB and BA domains occupy most of the bilayer's area, while AA nodes \cite{2022Cazeaux} are topological in nature \cite{paper1,paper2,paper3,Yoo2019,Xu2019,Moore,ordering,GorvachevReconstruction,BediakoReconstruction}, and their number cannot be modified once a relative angle of rotation has been set up. Molecular dynamics calculations were carried out at the highest allowed temperature in a molecular dynamics code to shed light on this question.

This study is based on molecular dynamics calculations, whose methodology is described in Sec.~\ref{sec:methods}. The results appear in Sec.~\ref{sec:results} and conclusions in Sec.~\ref{sec:conclusion}.

\begin{figure*}
\includegraphics[width=0.96\textwidth]{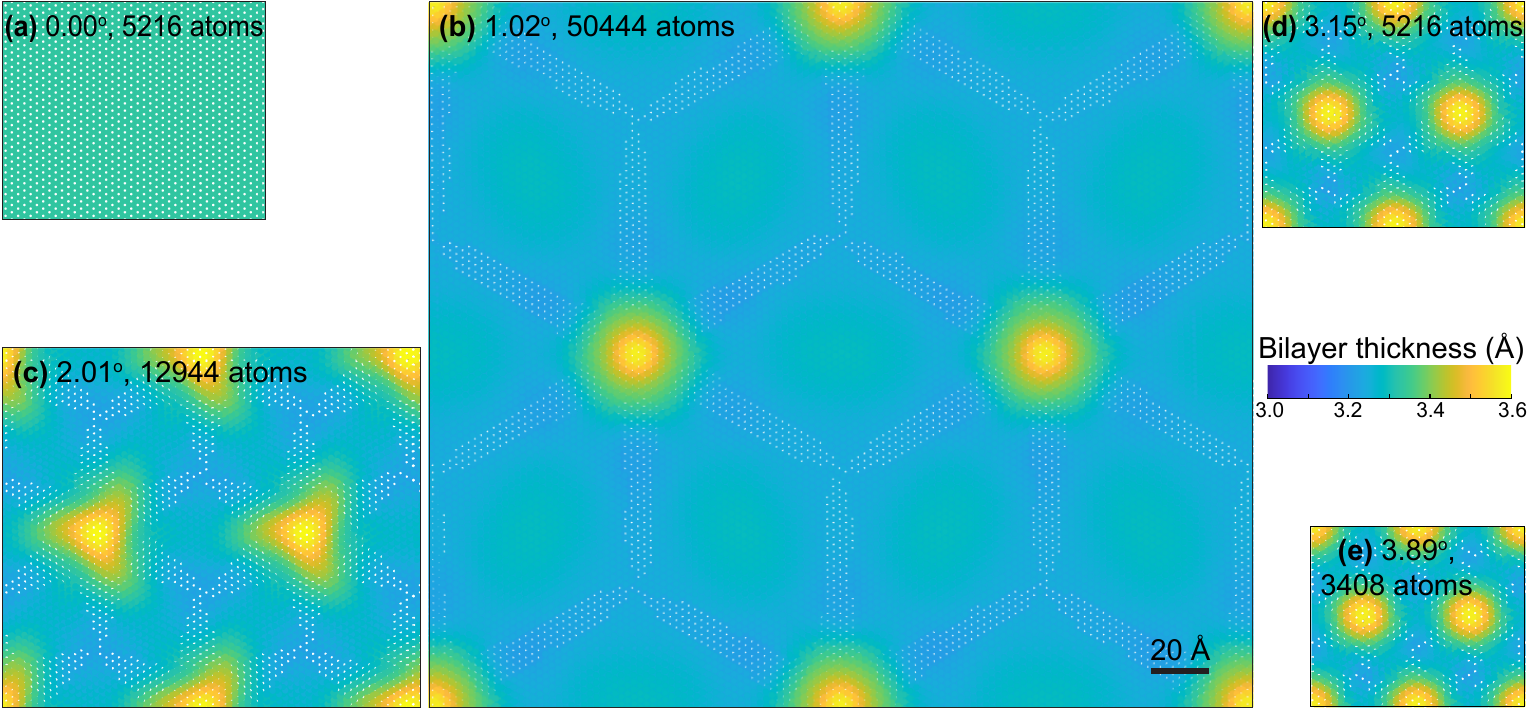}
\caption{Initial hBN bilayer structures after atomistic optimization at 1 K; false color indicates the relative height among layers, and the dimmest yellow locations indicate AA-nodes {(which are topological defects; see Ref.~\cite{2023Cazeaux})}. (a) A ferroelectric ($\delta=0^{\circ}$) sample {initially set into an AA relative configuration}. Moir\'e hBN bilayers with (b) $\delta=1.02^{\circ}$ and 50444 atoms, (c) $\delta=2.01^{\circ}$ and 12944 atoms, (d) $\delta=3.15^{\circ}$ and 5216 atoms, and (e) $\delta=3.89^{\circ}$ and 3408 atoms. All structures were drawn using the same length scale for a direct comparison.\label{fig:f_1}}
\end{figure*}

\section{Methods}\label{sec:methods}
%\subsection{Molecular dynamic simulations and interatomic potentials}
Atomistic models for hBN bilayers were constructed with an initial interatomic distance of 1.47 \AA{} and an interlayer distance of 3.33 \AA{}. The models consisted of commensurate periodic hBN bilayers with a relative rotation angle of $60^{\circ}-\delta$ degrees, with $\delta=0^{\circ}$ for a ferroelectric configuration, or $\delta=1.02^{\circ}$, $2.01^{\circ}$, $3.15^{\circ}$, and $3.89^{\circ}$, for moir\'e antipolar configurations. The studied structures contain 50,444, 12,944, 5,216, and 3,408 atoms, respectively. Two additional ferroelectric bilayers with either 200 or 400 atoms were also considered to better understand the drastically different sliding behaviors of ferroelectric and antipolar bilayers.

{Structures were initially set into an AA configuration, and forces were allowed to equilibrate in this position of (unstable) equilibrium in LAMMPS \cite{paper8}.} Molecular dynamics simulations were performed with the same numerical tool afterwards. Intr-alayer energetics were described by the Tersoff potential with parameters fitted for BN structures \cite{paper6,PhysRevB.84.085409}. {This interatomic potential accurately replicates the strain energy response, equilibrium lattice constant, and phonon dispersion relations of boron nitride nanostructures with data obtained from $x-$ray scattering experiments and density functional theory (DFT) calculations.} The anisotropic interlayer potential (ILP) for h-BN was utilized \cite{paper7} to represent the complex interlayer interactions. The velocity Verlet scheme was used for the time-integration of the resulting equations of motion with a constant time step of 0.001 ps. The interlayer energy $E_{int}$ was minimized until reaching a force threshold of $10^{-6}$  eV/\AA.  This permits elucidating AA nodes, which occur at the longest interlayer separation. An NPT equilibration stage (in which the number of atoms is fixed, pressure is set to 0 Pa, and temperature oscillate around 700 K) of $5 \times 10^{4}$ time steps was performed using the Nos\'e-Hoover thermostat \cite{paper9, paper10}.  Last, a production stage was set in an NVE ensemble (where the number of atoms, the volume, and total energy remain constant) for $5 \times 10^{6}$ time steps (5 ns) and a sampling rate of 0.5 ps.  Each monolayer’s center of mass' coordinates were recorded to detect their relative motion. {Additional molecular dynamics calculations were carried out at 300 K on a moir\'e bilayer with a $3.15^{\circ}$ relative rotation to determine the evolution of diffusion coefficients with temperature.}

\begin{figure}
\includegraphics[width=0.48\textwidth]{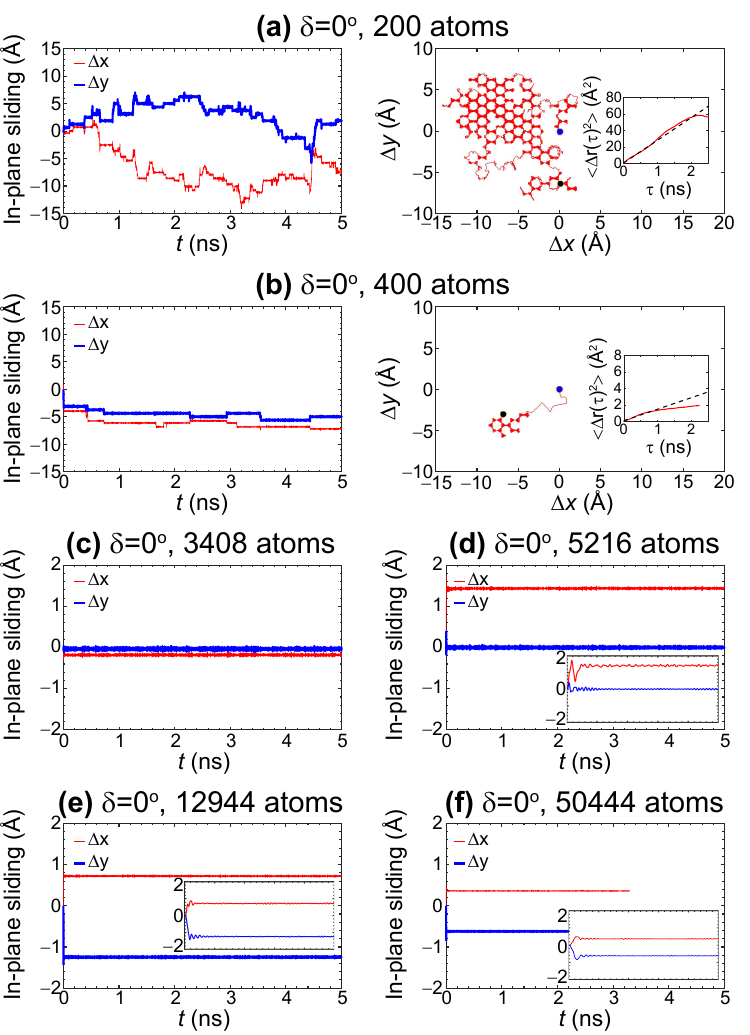}
\caption{Sliding of ferroelectric ($\delta=0^{\circ}$) hBN bilayers at 700 K {starting from an AA relative configuration}. Bilayers containing 200 and 400 atoms (subplots (a) and (b)) undergo telegraph-noise-like discrete displacements {$\Delta x$ and $\Delta y$} that sample the honeycomb lattice, similar to those described in Ref.~\cite{Marmolejo-Tejada2022}. The right panels contain the squared relative displacement of the monolayer's centers of mass  as a function of time $\langle\Delta r(\tau)\rangle^2$. As bilayers increase their size (subplots (c) through (f)), the single shear mode underpinning a whole concerted relative motion by one interatomic distance becomes suppressed, as seen by the straight lines and the comparatively smaller in-plane sliding scale. This means that a critical ferroelectric-to-paraelectric critical temperature cannot be uniquely assigned, as it depends on the bilayer's size. {In subplots (c) through (f), the bilayer quickly moves from the unstable AA configuration into a stable local minima (either AB or BA), and it does not move from that local minima for the extent of the simulations; inserts show the relative displacement of one monolayer as the bilayer falls into the energy minima.} \label{fig:f_2}}
\end{figure}

\section{Results}\label{sec:results}

\subsection{Size-dependent {sliding propensity}}

Structures optimized {in forces at zero temperature} are depicted in Fig.~\ref{fig:f_1}. Fig.~\ref{fig:f_1}(a) is a ferroelectric hBN bilayer of size similar to a moir\'e bilayer with $\delta=3.15^{\circ}$. The green color indicates a constant relative height among monolayers for the ferroelectric configuration. Figures \ref{fig:f_1}(b) through \ref{fig:f_1}(e) depict moir\'e bilayers with increasing $\delta$ (and a concomitant decrease in the number of atoms $n$). The regions in which individual constituent monolayers are farthest apart are colored in yellow. Those are energetically unfavorable AA-stacked nodes \cite{SrolovitzStackingEnergies,2022Cazeaux}.

{A precision concerning size effects and periodic boundary conditions is necessary: Even when under periodic boundary conditions, a system with $N$ atoms can only support 3N vibrational modes. Within the present context, the periodic boundary serves the goals of avoiding (i) dangling chemical bonds and (ii) large out-of-plane excursions commonly seen at the edges of (non-periodic) finite-size flakes.}

Figure \ref{fig:f_2} contains the first original result from this work: Refs.~\cite{Bauer2DFerroelectricsSliding} and us \cite{Marmolejo-Tejada2022} stated specific values for a critical temperature ($T_C$) on ferroelectric sliding bilayers by the onset of sliding events and{--starting from an unstable AA configuration--}Figs.~\ref{fig:f_2}(a) and \ref{fig:f_2}(b) do show discrete, telegraph-noise in-plane sliding {($\Delta x$,$\Delta y$)} events (see left panels in Figures \ref{fig:f_2}(a) and \ref{fig:f_2}(b)) consistent with discrete displacements resembling the honeycomb lattice  on molecular dynamics calculations at 700 K when the number of atoms is either 200 or 400 {(see the sampling of the honeycomb lattice on the right subplots in Figs.~\ref{fig:f_2}(a) and \ref{fig:f_2}(b))}. Nevertheless, the frequency of those steps decreases with the number of atoms (as do the relative displacements of the center of mass) until the frequency of those displacements turns into a complete halt (see the straight lines that imply no discrete jumps on Figs.~\ref{fig:f_2}(c) through \ref{fig:f_2}(f), where the number of atoms ranges from 3,408 to 50,444). This is to say that ferroelectric bilayers by sliding should have a {propensity to slide} that depends not just on an inherent energy barrier, but also on the sample's size. Similar size-dependent sliding events have been reported on graphene bilayers \cite{PhysRevB.101.054109}.

%%%%%%%%%%%%%%%%%%%%%%%%%%%%%
The observed phenomena can be explained as follows: assuming that barrier crossing can be described by one dimensional Smoluchowski equation~\cite{zwanzig2001nonequilibrium}, the rate of sliding events of ferroelectric bilayers $r_s$ could be written as:
\begin{equation}\label{eq:eq1}
r_s = \frac{D\beta\omega_{\rm min}\omega_{\rm max}}{2\pi}e^{-\beta\Delta U},
\end{equation}
where $D$ is the diffusion coefficient (to be determined momentarily), $\beta=1/k_BT$ is the inverse temperature, $\omega_{\rm min}$ and $\omega_{\rm max}$ are the frequencies at the minimum and the maximum of the potential well, and $\Delta U$ is the barrier height. The dependency of $D$ on the size of the bilayer, and on its ferroelectric or moir\'e antipolar phase, will be studied soon.

To quantify the relative sliding of monolayers seen in Figs.~\ref{fig:f_2}(a) and \ref{fig:f_2}(b), we calculated the square of the displacement of the atom nearest to the center of mass on one monolayer as a function of time $\langle\Delta r(\tau)\rangle^2$ defined as:
\begin{equation}
\langle\Delta r(\tau)\rangle^2 = \langle|\mathbf{r}_0(\tau+t)-\mathbf{r}_0(t)|\rangle^2
\end{equation}
(where $\mathbf{r}_0(t)$ is the relative sliding of the center of mass among monolayers at time $t$), and shown within insets in the right subpanels of Figs.~\ref{fig:f_2}(a) and \ref{fig:f_2}(b).

\subsection{Superlubric moir\'e bilayers}

\begin{figure}
\includegraphics[width=0.48\textwidth]{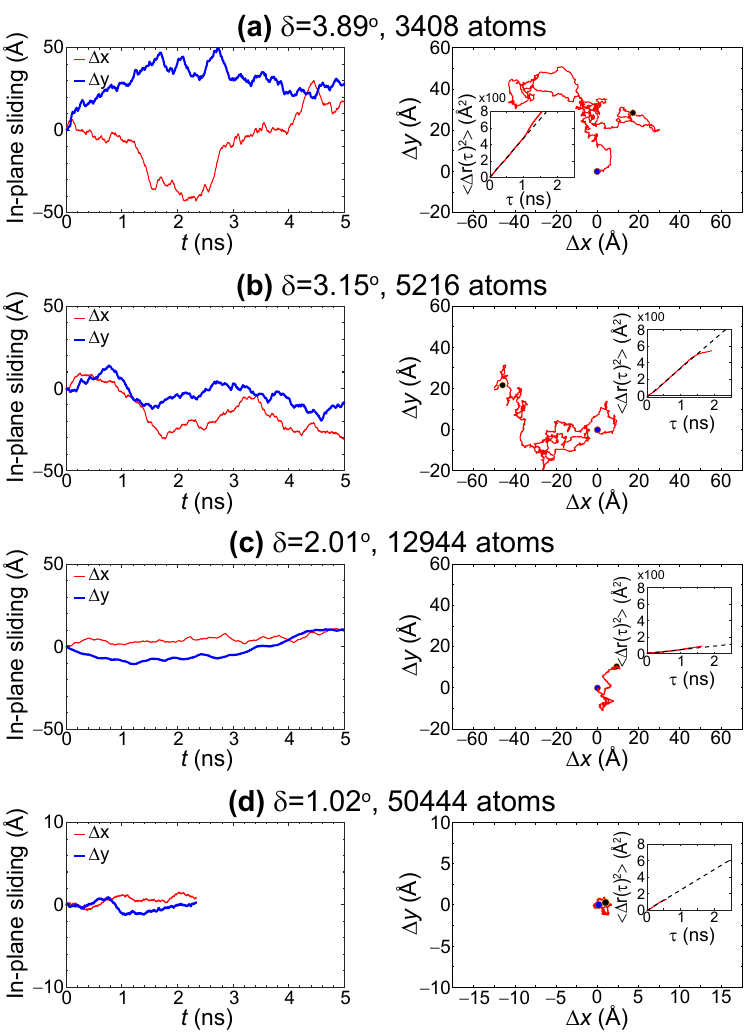}
\caption{Sliding of moir\'e hBN bilayers at 700 K. Similar to Figs.~\ref{fig:f_2}(a) and \ref{fig:f_2}(b), the left sub-panels show in-plane sliding displacements. The relative motion of the centers of mass of the individual monolayers does not track the honeycomb lattice anymore (compare right panels to those in Fig.~\ref{fig:f_2}(a) and \ref{fig:f_2}(b)), and the diffusion is orders of magnitude larger due to a lack of atomic registry upon relative rotation. Panels to the right contain $\langle\Delta r(\tau)\rangle^2$ as well and permit observing a suppressed diffusion as size increases.\label{fig:fig3}}
\end{figure}

The main difficulty in assembling moir\'es is the sudden, spontaneous unwanted relative motion of monolayers \cite{Lau} due to a loss of commensuration \cite{Hirano} upon rotation \cite{dienwiebel2004superlubricity}. A direct comparison among Figs.~\ref{fig:f_2} and \ref{fig:fig3} illustrates the different sliding behavior of ferroelectric and antipolar binary bilayers, which provides a mechanism to tell them apart.

To begin with, ferroelectric structures with 3408, 5216, 12944, and 50444 atoms have all commensurate unit cells containing four atoms which turn them ``rougher'' against relative thermally-induced displacements. This is the meaning of the straight lines on Fig.~\ref{fig:f_2}(c)-(f). The left subplots in Fig.~\ref{fig:fig3} display the in-plane sliding of the two monolayers in a moir\'e configuration. Even though they have the same number of atoms than the structures in Figs.~\ref{fig:f_2}(c)-(f), a relative sliding without discrete telegraph-like jumps can be clearly observed. Even more, the displacements occur over a range of 100 \AA, much larger than that seen in ferroelectric bilayers. The larger sliding of moir\'e bilayers with a lack of registry of the honeycomb lattice is evident in the right panels of Fig.~\ref{fig:fig3}, in which a chosen atom moves in what appears to be a continuous fashion.

\begin{figure}
\includegraphics[width=0.48\textwidth]{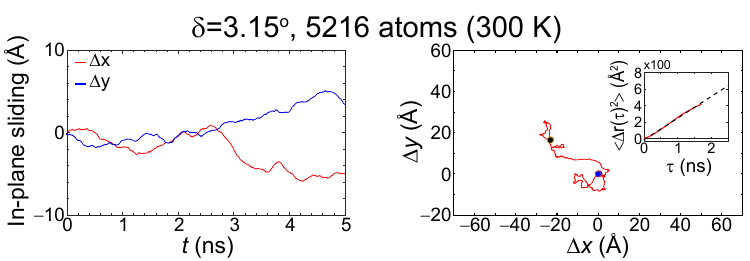}
\caption{Sliding of the moir\'e hBN bilayers with $\delta=3.15^{\circ}$ at 300 K. The left sub-panels show in-plane sliding displacements. The right panel contains $\langle\Delta r(\tau)\rangle^2$ and, when contrasted against Fig.~\ref{fig:fig3}(b), permit observing a suppressed diffusion as temperature decreases.\label{fig:fig4}}
\end{figure}

{We produced additional molecular dynamics calculations at 300 K for the bilayer at $3.15^{\circ}$ (containing 5216 atoms) and present those results in Fig.~\ref{fig:fig4}. Contrasted against the results shown in Fig.~\ref{fig:fig3}(b), the motion is restricted at the lower temperatures. Indeed, the diffusion coefficient $D$ is reduced from 0.0879 \AA$^2$/ps at 700 K, down to 0.0727 \AA$^2$/ps at 300 K.}

\begin{figure}
\includegraphics[width=0.48\textwidth]{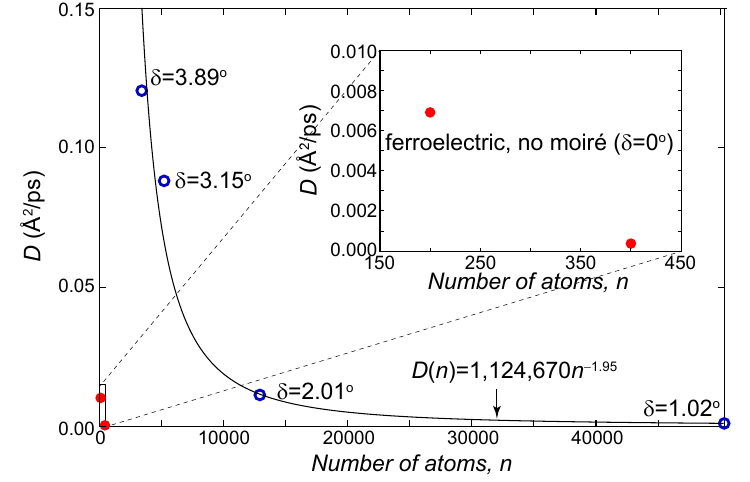}
\caption{Diffusion coefficient $D(n)$ obtained from the slope of $\langle\Delta r(\tau)\rangle^2=4D(n)\tau$. A power-law fit is included. The diffusion coefficients for two smaller ferroelectric bilayers are shown as well; note how they are orders of magnitude comparatively smaller for a given number of atoms.\label{fig:fig4}}
\end{figure}

The differences in sliding can be further ascertained by a comparison of the slopes of the $\langle\Delta r(\tau)\rangle^2$ {\em versus} $\tau$ plots, which depend on the number of atoms $n$. The idea is that, for diffusive dynamics, $\langle\Delta r(\tau)\rangle^2$ is related to the 2D diffusion constant $D(n)$ by:
\begin{equation}
\langle\Delta r(\tau)\rangle^2 = 4D(n)\tau.
\end{equation}
$D(n)$, plotted in Fig.~\ref{fig:fig4}, further confirms the striking difference in relative sliding among ferroelectric and antipolar bilayers. In connection to the ferroelectric-to-paraelectric transition in ferroelectric bilayers by sliding, one observes a quick decay of $D$ to zero as the number of atoms increases on the ferroelectric bilayer (inset on Fig.~\ref{fig:fig4}). The frequency of sliding effects $r_s$ is then shown to go to zero in Eqn.~\eqref{eq:eq1} along with $D(n)$, such that the critical temperature for a ferroelectric-to-paraelectric transition by sliding is size-dependent.

An analysis of friction of moir\'e bilayers is provided next.

\subsection{Friction within moir\'e bilayers}

We first calculate the coefficient of friction $\mu$ between the monolayers from our simulations (i.e., at $T=700$~K). $\mu$ is defined here as:
\begin{equation}
\mu = \left<\frac{F_f}{F_z}\right>,
\end{equation}
where $F_z\ge 0$ and $F_f$ are the instantaneous normal and friction forces, respectively, and $\left<\cdot\right>$ denotes a time average.
\begin{figure}[tb]
\begin{center}
\includegraphics[width=0.48\textwidth]{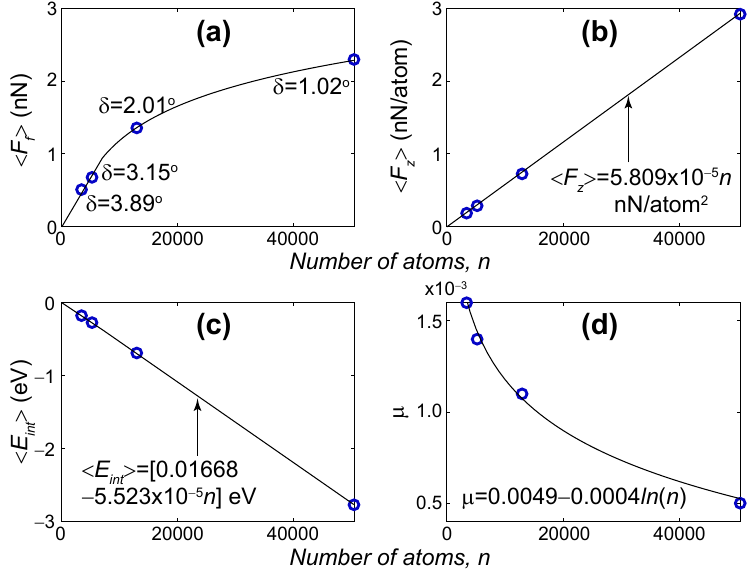}
\end{center}
\caption{Dependence of (a) average friction force $\left<F_f\right>$,  (b) average normal force $\left<F_N\right>$, (c) average interlayer energy $\left<E_{\rm int}\right>$ and (d) coefficient of friction $\mu$ on the number of atoms (i.e., on the angle $\delta$) for moir\'e hBN bilayers. The recorded values of $\mu$ imply that these are superlubric.}
\label{fig:friction01}
\end{figure}
The variation of $\left<F_z\right>$, $\left<F_f\right>$, average interlayer energy $\left<E_{\rm int}\right>$, and the coefficient of friction $\mu$ as a function of the number of atoms $n$ (and, implicitly, on the relative rotation angle $\delta$) for moir\'e hBN bilayers are depicted in Fig.~\ref{fig:friction01}. The friction and normal forces increase with $n$, but they follow different trends. The coefficient of friction remains in the superlubricity regime ($\mu<0.01$)~\cite{muser2014theoretical} for all values of $n$ ($\delta$). Rue {\em et al.}~determined $\mu=0.005$ for \ch{MoS2}/\ch{MoSe2} heterostructures~\cite{ru2020interlayer} at 300 K. for $\delta=5^{\circ}$ and $\langle F_z\rangle =0.2$~nN/atom. In our calculations at 700 K and for the largest angle we studied ($\delta=3.89^{\circ}$), $\langle F_z\rangle =0.0183$~nN/atom, and the coefficient of friction was $0.0016$, which is comparable with that of the \ch{MoS2}/\ch{MoSe2} heterostructure.

At this point, we use the Green-Kubo formalism to calculate the inter-layer friction constant $\lambda$. There, the interlayer friction constant $\lambda$ is defined as ~\cite{kubo1957statistical,kirkwood1946statistical,bocquet2013green,panoukidou2021comparison,shkulipa2005surface}:
\begin{equation}
\lambda = \lim_{t\to\infty} \frac{1}{2k_BTA}\int_{0}^t c(t')dt',
\end{equation}
where $k_B$ is the Boltzmann constant, $T$ is the temperature, $A$ is the contact area, and $c(t)$ is the lateral (in-plane) force correlation function defined as:
\begin{equation}
c(t)= \left<\mathbf{F}_{||}(t)\mathbf{F}_{||}(0)\right>.
\end{equation}

\begin{figure}
\begin{center}
\includegraphics[width=0.48\textwidth]{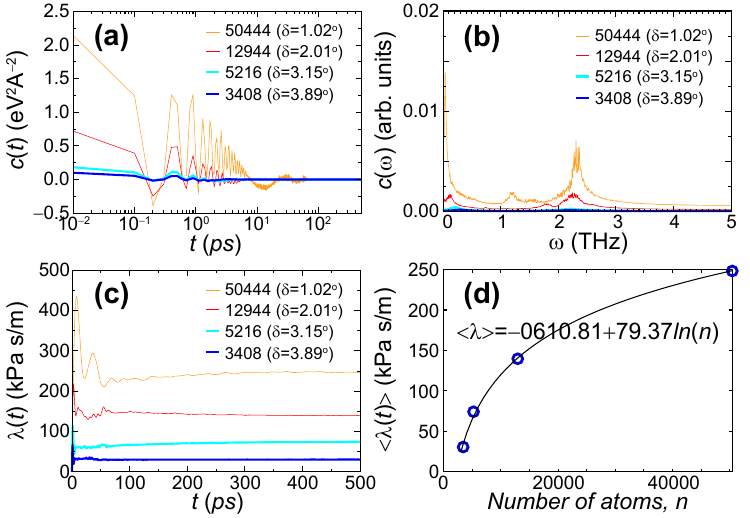}
\end{center}
\caption{(a) Time-autocorrelation of the lateral force between the monolayers. (b) Fourier transform of the force autocorrelation function. (c) Time dependence of $\lambda(t)$. (d) Friction constant $\lambda$ as a function of the number of atoms $n$.}
\label{fig:lambda}
\end{figure}

The force autocorrelation function $c(t)$ is shown in Fig.~\ref{fig:lambda}(a) for moir\'e hBN bilayers of different sizes ($\delta$) at $T=700$~K. For all values of $n$, $c(t)$ decreases rapidly and decays to zero at the time-scale of order of $200$~ps. In order to investigate the frequency response of the friction constant, we next looked the Fourier transform of $c(t)$:
\begin{equation}
c(\omega) = \int_0^{\infty} c(t) e^{-i\omega t} dt.
\end{equation}
$c(\omega)$ was calculated for different values of $\delta$ in Fig.~\ref{fig:lambda}(b). We find that $c(\omega)$ exhibits two peaks--one at lower frequency ($\approx 0.1$~THz), and another at slightly higher frequency ($\approx 2.0$~THz). The amplitude of the lower frequency peak becomes smaller and moves to slightly higher frequency, while the amplitude of the high-frequency peak becomes weaker  upon decreasing $n$. In Fig.~\ref{fig:lambda}(c), we show the behavior of $\lambda(t)$ for four different values of $n$. The values of the friction constant obtained from the long-time behavior of $\lambda(t)$ are shown as a function of $n$ in Fig.~\ref{fig:lambda}(d);  $\lambda$ increases monotonically with increasing $n$. The behavior of the friction constant is consistent with the increasing value of friction force with $n$ discussed in Fig.~\ref{fig:friction01}(a).

\begin{figure*}[tb]
\includegraphics[width=0.96\textwidth]{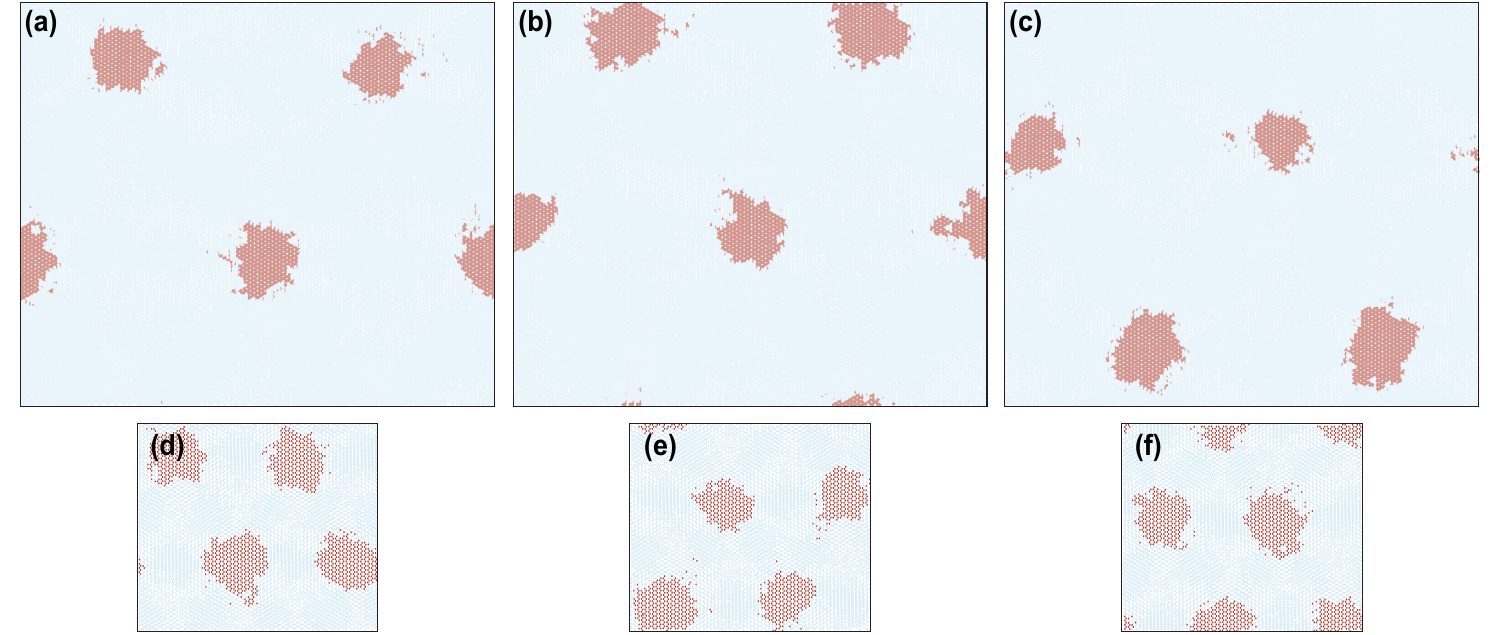}
\caption{The number of topological defects (dark-colored patches corresponding to AA-stacked sections \cite{2023Cazeaux,2022Cazeaux}) for the moir\'e hBN-bilayer with $\delta = 1.02^{\circ}$ at (a) 2.05 ns, (b) 2.15 ns, and (c) 2.30 ns and at 700 K. Subplots (d) through (f) display the moir\'e bilayer at 700 K at $2.01^{\circ}$ at 4.295 ns, 4.345 ns, and 4.395 ns, respectively. In both cases, a diameter of about 50 \AA{} can be observed for the AA-nodes.\label{fig:fig7}}
\end{figure*}

\subsection{Observations on an antipolar to paraelectric phase transitions}
{The network of AB and BA domains forms a network separated by dislocation lines meeting at AA nodes \cite{2023Cazeaux,2022Cazeaux}. Engelke and coauthors posited that one can map the local relative sliding configurations in individual unit cells though the moir\'e into a closed space akin to a punctured torus, where the punctured section is precisely the AA-node. In this map, the torus is the closed surface in which the winding of a local displacement field turns out to be topological \cite{2023Cazeaux}.} And so, the last item for discussion is the possibility of a mechanism by which the density of topologically-protected AA-nodes alluded on the previous paragraph \cite{2023Cazeaux,2022Cazeaux} could be thermally altered. That possibility was the reason to perform molecular dynamics calculations at 700 K in the first place (which was the highest we could achieve without facing numerical instabilities in the code).

Figure \ref{fig:fig7} represents a qualitative answer to the question. {Figures \ref{fig:fig7}(a-c) show} the relative height among the two monolayers in the {$\delta=1.02^{\circ}$} configuration at three different times (2.05 ns,  2.15 ns, and 2.30 ns) along the molecular dynamics evolution. The darkest hue is a 3.5 \AA{} cutoff, above which the local bilayer has a local AA-stacking. The point is that, though those move around, {\em four} AA-stacked sections always remain visible on the snapshots.

{We expect the antipolar behavior described by Bennett et al.~\cite{Bennett3} to still be present at high temperature. Then, we showed by molecular dynamics calculations that the AA-nodes cannot be thermally removed. Therefore, to produce a transition onto a ferroelectric or paraelectric phase, the only options are either (i) a quantum phase transition, or (ii) a topological one. As the temperature is high, a quantum phase transition is discarded as well, and a topological transition signified by the removal of AA-nodes is the only possibility left. Indeed, the rotation by $-\delta$ renders the antipolar moir\'e onto a ferroelectric configuration which may be unstable to sliding events depending on the samples' size and temperature.}

\section{Conclusion}\label{sec:conclusion}

%%%%%%%%%%%%%%%%%%%%%%%%%%%%%%%%%%%%%%%%%%%%%%%%%%%%%%%%%%%%%%
To conclude, we undertook a study of hBN bilayers in the ferroelectric ($\delta=0^{\circ}$) and moir\'e antipolar ($\delta >0^{\circ}$) configurations. We first established the fact that ferroelectric-to-paraelectric transitions by sliding occur at temperatures dependent on system size. Then, using configurations with identical number of atoms, we establish an easier propensity to slide of antipolar moir\'e bilayers. This study included the calculation of the diffusion coefficient, and of parameters $\mu$ and $\lambda$ that indicate superlubric behavior. The work ends by the observation that the area density of topological defects (given by AA-nodes) remains protected at the highest temperature of 700 K, in which molecular dynamics simulations could run without developing numerical instabilities. {Therefore, the only possibility for the removal of AA-nodes is a rotation out of the topologically-nontrivial moir\'e conformation onto a (trivial) configuration with a single realization of relative sliding among monolayers across all unit cells.}

%%%%%%%%%%%%%%%%%%%%%%%%%%%%%%%%%%%%%%%%%%%%%%%%%%%%%%%%%%%%%%

\section{Acknowledgments}
S.B.-L.~acknowledges funding from the U.S.~Department of Energy (Award DE-SC0022120).

%\bibliography{ref}
%\bibliographystyle{apsrev}

\providecommand{\noopsort}[1]{}\providecommand{\singleletter}[1]{#1}%

\end{document}